\documentclass[12pt]{article}

\usepackage{amsmath,amsthm,amsfonts, latexsym}
\usepackage{graphicx}

\newcommand{\mtx}[2]{\left(\begin{array}{#1}#2\end{array}\right)}

\begin{document}

\begin{center}

\bigskip
{\Large Distinguishing unentangled states with an unentangled measurement}\\

\bigskip

William K.~Wootters \\

\bigskip

{\small{\sl

Department of Physics, Williams College, Williamstown, 
MA 01267 }}\vspace{3cm}

\end{center}

\subsection*{\centering Abstract}
{In a 1991 paper, Asher Peres and the author theoretically analyzed a set of unentangled bipartite
quantum states that could apparently be distinguished better by a global measurement
than by any sequence of local measurements on the individual subsystems.  The present
paper returns to the same example, and shows that the best result so far achieved can alternatively 
be attained by a measurement that, while still global, is ``unentangled'' 
in the sense that the operator associated with each measurement outcome is a tensor product. }

\vfill

%PACS numbers: 03.65.Ta, 02.10.-v

\newpage

\section{Introduction}

In the fall of 1989, I had the good fortune to be at the Santa Fe Institute at the same time as Asher
Peres, and we quickly began working together on the following problem.

\begin{quote}
A quantum system consists of two spatially separated components.  The system
is known to be in one of several possible pure states, each
of which is a {\em product} state, and one wants to perform a measurement that will 
provide as much information as possible about the identity of the system's state.  
One can consider two kinds of measurement: (i) a sequence of local measurements
on the individual subsystems, each measurement possibly depending on 
the outcomes of earlier 
measurements; (ii) a single global measurement on the system as a whole.  
{\em Question}:  In order to gain as much information as possible, is it sometimes
necessary to use a global measurement, or can one always do just as well with
a cleverly chosen sequence of local measurements?  
\end{quote}
The states that one is trying to distinguish are all product states;
so one might have expected that local measurements
would be sufficient.  Indeed, this was my intuition at the time.  Fortunately it contrasted nicely 
with Asher's intuition, and the contrast helped fuel a stimulating collaboration.
At first we considered a few examples in which one was trying to distinguish
just {\em two} states; in these cases we found no advantage
to a global measurement.  (A few years later, Ban {\em et al.}~\cite{BYH}, following work by
Brody and Meister \cite{BM}, showed that for distinguishing between just two 
states $|\psi_1\rangle\otimes|\psi_1\rangle$ and $|\psi_2\rangle\otimes|\psi_2\rangle$, the optimal
amount of information can always be obtained by a sequence of local measurements.
Recently a similar conclusion was reached, for an arbitrary number of copies, 
by Acin {\em et al.}~\cite{ABBMM}.)
Finally, though, we hit upon
a three-state example for which a global measurement was distinctly better than
any adaptive local strategy that we could devise \cite{PW}.
Though the superiority of global measurements in this particular example has still not been proved
conclusively,
other studies, based on other examples, have decisively
confirmed that Asher's intuition about global measurements was correct
\cite{B,UPB1,UPB2}.

In the present paper I return to the same three-state example
that we originally considered---the ``double-trine'' ensemble---in order to explore a further distinction among measurements.  In 1989 Asher convinced me (and I remain convinced)
that no sequence of local
measurements can match the best global measurement in discriminating among 
the three double-trine states.
But among all complete, global measurements, it is interesting to consider
two mutually exclusive and exhaustive classes: (i) those whose outcomes
are associated with product states, and (ii) those for which at 
least one outcome is associated with an entangled 
state.   I will call the former measurements ``unentangled'' and the 
latter ``entangled.''\footnote{Some authors, including the present author,
have referred to measurements of the former class as ``separable measurements,''
since they can be implemented by separable superoperators \cite{B,UPB1,UPB2}.
But other authors have reserved the term ``separable'' for sequences of 
local measurements (see, for example, Refs.~\cite{GM, S, BBM}).  In this paper
I am using the term ``unentangled measurement'' in order to avoid confusion.}
One might have thought that any measurement whose outcomes are associated 
with product states could be carried out as a sequence of local measurements,
but this is not the case \cite{B}.  That is, the class of unentangled measurements
is strictly larger than the class of measurements that can be realized locally, 
even with classical communication. 
Still, unentangled measurements are quite special.  For example, like local
adaptive measurements, they do not create
entanglement where none existed previously.  And as a class they are more tractable
mathematically than sequences of local measurements.  So it is interesting to 
ask how effective such measurements can be, and to what extent they are limited relative
to the much larger class of entangled measurements.  

For the double-trine ensemble, the best measurement that Asher and I found was
not merely a global measurement but also an entangled measurement.  So our
result left
open the question whether an unentangled measurement could do just as well.  
The main purpose of the present paper is to show that an unentangled measurement
{\em can} in fact do just as well, as we will see in Section 3.

Though this paper focuses on a specific example, it raises a more general question.
We know by now several examples of ensembles of product states for which the optimal
discrimination cannot be achieved by a sequence of local 
measurements (see, for example, Refs.~\cite{B,UPB1,UPB2}).  However,
I have not found in the literature any example for which it is known that the optimal
discrimination cannot be achieved by an {\em unentangled} measurement.  The case
that Asher and I considered might have seemed a good candidate for such an example,
since the best measurement that we found is entangled.  But the
result of the present paper rules out this case.  One wonders, then, whether {\em every}
ensemble of product states can be optimally distinguished by an unentangled measurement.
Though this may seem unlikely, it has apparently not been shown to be false.  

Asher returned to Israel before we had finished working on the problem, so we
continued our collaboration by email.  I still have hard copies of most if not all of the 
notes that I received from him during those months; most of these advance the research itself 
but some address other matters.  In one of these emails, Asher responds to a very early draft
of our paper that I had sent him, in which I had used the word ``considerably,''
and at one point, also the word ``mind,'' as in ``the particles' correlation is only in the 
mind of the observer.'' He politely noted that he would prefer not to use the word
``mind'' in this context---he said that everyone is free to interpret the formalism
according to one's cultural background---and that in general he preferred to avoid the words
``extremely'' and ``considerably,'' a practice he learned from Larry Schulman
and later found supported in {\em The Elements of Style} by Strunk and White.  
I confess that I did not learn this lesson and have allowed ``considerably'' to slip into
many papers since then. But in honor of Asher, and for the sake of good
style, I will resist the temptation to use either ``considerably'' or ``extremely'' 
in the rest of this paper.  There should be no need for ``mind'' either.

\section{The double-trine ensemble}

The ensemble considered in Ref.~\cite{PW} consists of three pure product states of a pair of
qubits, each state having {\em a priori} probability 1/3.  
Each qubit is in one of three states equally spaced on a great circle of the Bloch
sphere, and
the two qubits are both in the {\em same} state.  
We represent the three single-qubit 
states---called the ``trine'' states---as
\begin{equation}
|\psi_0\rangle = \mtx{c}{1 \\ 0} \hspace{8mm}
|\psi_1\rangle = \mtx{c}{-1/2 \\ -\sqrt{3}/2} \hspace{8mm}
|\psi_2\rangle = \mtx{c}{-1/2 \\ \sqrt{3}/2}
\end{equation}
Here the overall phases of the vectors have been chosen so that
$\langle\psi_i|\psi_j\rangle$ has the same value, $-1/2$, for all $i\neq j$.  
(This choice makes some later equations
simpler.)  
%The states $|\psi_j\rangle$ are shown in Fig.~1.  

%\begin{figure}[h]
%\hfill
%\includegraphics[scale=0.8]{newtrine.pdf}
%\hfill
%\smallskip
%\caption{The three trine states.  The figure shows the $x$-$z$ plane of the Bloch sphere,
%in which the trine states lie.}
%\end{figure}

The three possible states of the {\em pair} of qubits are
\begin{equation}
|a_j\rangle = |\psi_j\rangle \otimes |\psi_j\rangle, 
\hspace{7mm}
j = 0, 1, 2.
\end{equation}
(I will use Latin letters for two-qubit states and Greek letters
for single-qubit states.)  
We imagine that we are presented with a single copy of 
such a pair of qubits; our goal is to perform a measurement
on the pair that distinguishes the three states as well as
possible.   

As our measure of success, we use the mutual information between
the outcome of our measurement and the index $j$ that identifies
the state ($j= 0,1,2$).  Let the outcome of the measurement be
labeled by $k = 1, \ldots, M$, where the number of outcomes, $M$,
need not be the same as the 
number of states in our ensemble.  The mutual information $I$ is
\begin{equation}
I = H(\hbox{state}) + H(\hbox{outcome}) - H(\hbox{state and outcome}),
\end{equation}
where $H$ is the Shannon entropy of the indicated probability
distribution.  Let $p_k$ be the probability of outcome $k$ (averaged over
the possible states of the system), and 
let $p_{jk}$ be the probability that the state is $|a_j\rangle$
and the measurement outcome is $k$.  Then for our three-state
example we can write the mutual
information as
\begin{equation}
I = \log 3 - \sum_{k=1}^M p_k \log p_k
+ \sum_{j=0}^2 \sum_{k=1}^M p_{jk} \log p_{jk},
\end{equation}
in which we take the base of the logarithms to be 2.  For our purposes,
a somewhat more convenient form is 
\begin{equation}
I = -\sum_{k=1}^M p_k\log p_k+ (1/3)\sum_{j=0}^2\sum_{k=1}^M p_{k|j}
\log p_{k|j},  \label{mutinf}
\end{equation}
where $p_{k|j}$ is the probability of outcome $k$ given that the
state is $|a_j\rangle$.  In this form, it reads as the average
amount of information one gains about the outcome of the measurement
upon learning the identity of the state, but the mutual information
can also be interpreted as 
the average amount of information that one gains about the state
upon seeing the outcome of the measurement.  Note that with three
equally likely states in the ensemble, 
the maximum conceivable value of the mutual information
is $\log 3 = 1.585$ bits, the value we would get if the measurement
outcome were perfectly correlated to the system's state.  We will
certainly not achieve this value, since our three states are not
mutually orthogonal and can therefore not be distinguished perfectly.

In Ref.~\cite{PW} a number of different measurement strategies for
this problem were considered, the best being an orthogonal joint
measurement on the pair of qubits.  The construction of this measurement
is motivated by the observation that the three states $|a_0\rangle$,
$|a_1\rangle$, and $|a_2\rangle$ can be regarded as unit vectors
in a real vector space, the angle between each pair being $75.5^\circ$.  
(This angle is the inverse cosine of 1/4, which
is the inner product between any two of these vectors.)  There is
a unique triple of {\em orthogonal} vectors in this real vector
space that symmetrically 
straddles the vectors $|a_0\rangle$,
$|a_1\rangle$, and $|a_2\rangle$ and approximates them as closely 
as possible.  One finds that the three
orthogonal vectors are 
\begin{equation}
|A_j\rangle = \frac{1}{3\sqrt{3}}\big[(4+\sqrt{2})|a_j\rangle
- (2 - \sqrt{2})(|a_{j+1}\rangle + |a_{j+2}\rangle ) \big], \hspace{3mm}
j = 0,1,2,
\end{equation}
where the addition in the subscripts is mod 3.  To complete the
measurement on the pair of qubits, we need a fourth state orthogonal
to each of the states $|A_j\rangle$.  This is the singlet state
\begin{equation}
|S\rangle = \frac{1}{\sqrt{2}}(|01\rangle - |10\rangle),
\end{equation}
where $|0\rangle$ and $|1\rangle$ are our standard basis states
for a single qubit.  One can use the relation $\langle a_i|a_j\rangle
= (1/4)+ (3/4)\delta_{ij}$ to
verify that the states
$|A_j\rangle$ are indeed mutually orthogonal.  That they are also all orthogonal
to the singlet state follows from the construction
of $|a_j\rangle$ as a
repeated single-qubit state, which places it in the triplet subspace.
This latter orthogonality implies that for the given ensemble,
the outcome $|S\rangle$ will never happen.  So for our purpose this
measurement has, in effect, only three possible outcomes.    

One can show that the three states $|A_j\rangle$ are all entangled,
as is the singlet state $|S\rangle$.  So the 
measurement defined by these states is an {\em entangled}
measurement.  

How large a value of the mutual information does this measurement
provide?  To answer this question we need to compute the probability
of the outcome $|A_k\rangle$ given the initial state $|a_j\rangle$.
For the case $k=j$, we have
\begin{equation}
p_{j|j} = |\langle a_j|A_j\rangle |^2
= \left\{\frac{1}{3\sqrt{3}}\left[ (4+\sqrt{2}) - (1/2)(2-\sqrt{2})\right]
\right\}^2 = \frac{1}{2}+\frac{\sqrt{2}}{3}, \label{whole}
\end{equation}
which is about 0.971.  (In the real-vector-space picture described above,
the angle between each vector $|a_j\rangle$ and its corresponding
$|A_j\rangle$ is $\cos^{-1}(\sqrt{0.971}) \approx 10^\circ$.)
The other two outcomes share the 
remaining probability equally; so for $k\neq j$ we have
\begin{equation}
p_{k|j} = \frac{1}{2}\left[ 1 - \left( \frac{1}{2}+\frac{\sqrt{2}}{3}
\right) \right] = \frac{1}{4} - \frac{1}{3\sqrt{2}}, 
\end{equation}
which is about 0.014. 
The three measurement
outcomes have equal {\em a priori} probability; so according to 
Eq.~(\ref{mutinf}), the mutual information is
\begin{equation}
I = \log 3 + (0.971)\log(0.971) + 2(0.014)\log(0.014) 
= 1.369\;\hbox{bits},
\end{equation}
which is not much less than the upper bound $\log 3 = 1.585$ bits.
Because of the symmetry of the problem and the proximity of
each outcome vector $|A_j\rangle$ to the corresponding
state vector $|a_j\rangle$, it is plausible that the measurement
we have just described is optimal.  Numerical work similarly suggests
that it is optimal \cite{Shor}, but this
conjecture has apparently never been proved, and I have no proof of it to 
offer here.  (Refs.~\cite{Ban, Sasaki, EF} prove that it is optimal in different senses,
that is, with measures of success other than the mutual information.)

The other measurement strategies considered in Ref.~\cite{PW} are all
sequences of local measurements, the most sophisticated being a
back-and-forth sequence of successively stronger measurements,
alternating between the two qubits.  But the highest value of
mutual information obtained in this way was $1.26205 \pm 0.00037$
(the uncertainty was computed from a Monte Carlo simulation),
and it seems unlikely that any such sequence will match the
value 1.369 obtained from a global measurement.  

As far as I know, this last conjecture has also never been proved.
However, the claim that global measurements can be better,
even when the states in question
are product states, has been demonstrated decisively
in other contexts, as I have noted in the Introduction.  Bennett {\em et al.} have given an example of
a set of {\em orthogonal} product states that cannot be
distinguished by any sequence of local measurements,
even though, being orthogonal, they certainly can be distinguished
by a global measurement \cite{B}.   Similar examples have been given
as part of the study of ``unextendible product bases'' \cite{UPB1,UPB2},
and efficient proofs of this sort of indistinguishability have been 
devised \cite{GV,WH}.   There is also an extensive literature
on the problem of ascertaining a quantum state, or some parameter
of a quantum state, when many instances of the state
are provided; many of these papers distinguish between local and global
measurements and ask whether the latter 
allows a more precise estimate (see, for example, Refs.~\cite{MP, GM,
S, ABBMM, BBM} and references cited therein).  One finds that the existence 
of an advantage conferred by global measurements
depends on the precise question being asked.

In the following section I continue analyzing the double-trine
ensemble, asking in particular how well the three states can
be distinguished by an {\em unentangled} measurement.

\section{An unentangled measurement}

Consider a {\em single} qubit, known to be in one of the three
trine states $|\psi_j\rangle$, $j = 0,1,2$, with equal {\em a priori}
probabilities.  It has been shown\footnote{This three-outcome
POVM was proposed for this ensemble by Holevo \cite{H}, who showed that it
provided more information than any orthogonal measurement.
That this POVM is in fact optimal was proved
by Sasaki {\em et al.}~\cite{SBJOH}.} that the optimal mutual information
for this case can be obtained by a three-outcome, non-orthogonal 
measurement represented by the following positive-operator-valued
measure (POVM):
\begin{equation}
\Pi_k = \frac{2}{3}|\psi_k^\perp\rangle \langle \psi_k^\perp |, 
\hspace{5mm} k = 0, 1, 2.  \label{1qubit}
\end{equation}
Here the operators $\Pi_k$ are the positive operators representing 
the outcomes of the measurement---their sum is the $2\times 2$ identity---and
$|\psi_k^\perp\rangle$ is the qubit state orthogonal
to $|\psi_k\rangle$.  The probability of outcome $k$ if the qubit
is in the state $|\psi_j\rangle$ is
\begin{equation}
p_{k|j} = \langle\psi_j|\Pi_k |\psi_j\rangle = 
\left\{\begin{array}{l}
0, \;\hbox{if $k=j$} \\
1/2, \; \hbox{if $k=j+1$} \\
1/2, \; \hbox{if $k=j+2$}
\end{array}
\right.
\end{equation}
where the addition is mod 3.
Thus this measurement has the effect of ruling
out one of the three possible states and leaving the other two
equally likely.  The mutual information is $I = \log 3 - \log 2
= 0.585$ bits, which is strictly larger than any value that can be
obtained with an orthogonal measurement.  

Let us try to use the idea behind this measurement to guide us in 
designing an unentangled measurement for the double-trine ensemble.
As indicated in the Introduction, by ``unentangled'' I mean that
each of the operators of the POVM will be a tensor product.    
Now, initially we have three possible states of the pair of qubits.
Ideally, we would like to rule out two
of these states by our measurement.  We have just seen an example of
a single-qubit measurement that rules out {\em one} of the three.
Since we now have {\em two} qubits to work with, we can try to use them
to rule out {\em two} states, thereby leaving us with
only the correct state.  (We know in advance that this will not
in fact be possible, since the three states of our ensemble are
not orthogonal, but let us see how well we can do.)  
As our first guess towards a good POVM, consider the following set:
\begin{equation}
\left\{ \Pi_0\otimes\Pi_1,\; \Pi_1\otimes\Pi_0,\;
\Pi_1\otimes\Pi_2, \;\Pi_2\otimes\Pi_1,\;
\Pi_2\otimes\Pi_0,\; \Pi_0\otimes\Pi_2\right\}, \label{bad}
\end{equation}
where each of the operators $\Pi_j$ is chosen from
the single-qubit POVM given in Eq.~(\ref{1qubit}).  The reader may
already have noticed that this set does not actually constitute
a POVM, because its elements do not add up to the 
$4\times 4$ identity operator.  Ignoring for now this annoying fact, we
note that any of the six operators listed here
{\em would be} ideal as an element of a POVM, since each rules out
two of the states of our ensemble.  For example, 
$\langle a_j|\Pi_0 \otimes \Pi_1|a_j\rangle$ is nonzero
only if $j=2$.  

Can we modify slightly the set given in Eq.~(\ref{bad})
so as to create a legitimate POVM whose effect
approximates the ideal?  One approach would be to
include three additional elements: $\Pi_0\otimes \Pi_0$,
$\Pi_1\otimes \Pi_1$, and $\Pi_2\otimes \Pi_2$.  
Then one is in effect performing on each of the qubits
the POVM given in Eq.~(\ref{1qubit}), and this is a perfectly
legitimate measurement on the pair.  It turns out, though, that
this is not a very effective strategy: the mutual information
is $\log 3 - 1/2 = 1.085$ bits, which is not nearly as large
as for the entangled measurement considered in Section 2.\footnote{Though
this particular strategy is not mentioned in Ref.~\cite{PW}, on looking
back over the email correspondence, I see that Asher did 
consider this possibility.  In a note dated
5 Nov 1989, he observed that even though this method does not require classical communication
between the two parties, the amount of information it provides, 1.085 bits, is larger than
what can be obtained with a pair of local {\em orthogonal} measurements
connected by classical communication.}
(The reason
for the poor discrimination is 
that these last three elements, each of which rules out
only one of the three states, have a combined probability
equal to the combined probability of the six ``good'' POVM elements
listed in Eq.~(\ref{bad}).)

A better strategy is to construct a POVM with only six elements
by modifying slightly each of the operators in Eq.~(\ref{bad}).  Rather than insisting
that each POVM element definitively rule out two of the states
$|\psi_j\rangle$, we look for POVM elements that will make two of
the states very unlikely.  To this end, we define six single-qubit
states as follows:
\begin{equation}
|\phi_j^{+}\rangle = R|\psi_j\rangle, \hspace{5mm}
|\phi_j^{-}\rangle = R^{-1}|\psi_j\rangle, \hspace{5mm}
j = 0, 1, 2.
\end{equation}
Here the unitary matrix $R$ is given by
\begin{equation}
R = \mtx{cc}{\cos(\theta/2) & -\sin(\theta/2)
\\ \sin(\theta/2) & \cos(\theta/2)},
\end{equation}
where the angle $\theta$ is yet to be determined.  
In the $x$-$z$ plane of the Bloch sphere, the states
$|\phi_j^+\rangle$ and $|\phi_j^-\rangle$ each make
an angle $\theta$ with $|\psi_j\rangle$ and lie on
opposite sides of it.
% as shown in Fig.~2.  

%\begin{figure}[h]
%\hfill
%\includegraphics[scale=0.8]{newtrineplus.pdf}
%\hfill
%\smallskip
%\caption{The six states $|\phi_j^\pm\rangle$ from which we seek to construct an unentangled 
%measurement.}
%\end{figure}

We now consider the following
six two-qubit states:
\begin{equation}
|B_j\rangle = |\phi_j^+\rangle\otimes |\phi_j^-\rangle,
\hspace{5mm}
|C_j\rangle = |\phi_j^-\rangle\otimes |\phi_j^+\rangle,
\hspace{5mm}
j = 0, 1, 2.
\end{equation}
Our new candidate for a good POVM consists of the operators
\begin{equation}
E_j = \alpha |B_j\rangle\langle B_j|, \hspace{8mm}
F_j = \alpha |C_j\rangle\langle C_j|, \hspace{5mm}
j = 0, 1, 2,
\end{equation}
where $\alpha$ is a constant whose value is to be determined.
We want to choose $\theta$ and $\alpha$ so that the set
${\mathcal E} = \{E_1, E_2, E_3, F_1, F_2, F_3\}$ constitutes a POVM.

Notice that if $\alpha = 4/9$ and $\theta = 60^\circ$, the set
${\mathcal E}$ is identical to the set given in Eq.~(\ref{bad}).
In that case, for example, $|\phi_0^+\rangle$ equals  
$|\psi_2^\perp\rangle$ and $|\phi_0^-\rangle$ equals $|\psi_1^\perp\rangle$,
so that $E_0$ becomes proportional to $\Pi_2 \otimes \Pi_1$.
But this choice does not give us a POVM.  For arbitrary 
$\alpha$ and $\theta$, one can work out the sum of the elements
of ${\mathcal E}$; the result is
\begin{equation}
\sum_{j=0}^2 (E_j + F_j) = \alpha\left(\frac{3}{2}\right)
\left[I + \frac{1}{2}\cos(2\theta)(\sigma_x\otimes\sigma_x + \sigma_z\otimes \sigma_z)
\right],
\end{equation}
where $\sigma_x$ and $\sigma_z$ are Pauli matrices.  This sum is 
equal to the identity only if $\alpha = 2/3$ and 
$\cos(2\theta)= 0$.  (That $\alpha$ must equal 2/3 can in fact be
derived more easily just by insisting that the 
{\em traces} of the elements of ${\mathcal E}$ add up to the 
trace of the identity.)  We thus obtain a POVM by choosing $\theta$
to be either $45^\circ$ or $135^\circ$.  The choice $\theta = 45^\circ$
is closer to our ideal of $60^\circ$, and indeed one finds that
this choice provides a significantly larger mutual information.  

With this choice---$\alpha = 2/3$ and $\theta = \pi/4$---let us now
compute the mutual information.  Because of the symmetry, it is sufficient
to find the probability of, say, the outcome $E_0$ when the state
is $|a_0\rangle$; all other probabilities can be computed from this one.
This probability is
\begin{equation}
\hbox{(probability of $E_0$ given $|a_0\rangle$)}
= \langle a_0|E_0|a_0\rangle
= \frac{2}{3}|\langle \psi_0|\phi_0^{+}\rangle|^2
|\langle \psi_0|\phi_0^{-}\rangle|^2.
\end{equation}
Now, the state $|\phi_0^+\rangle$ is $45^\circ$ away from
$|\psi_0\rangle$ on the Bloch sphere, and so is $|\phi_0^-\rangle$.
So $|\langle \psi_0|\phi_0^{+}\rangle|^2
= |\langle \psi_0|\phi_0^{-}\rangle|^2 = \cos^2(\pi/8)$.  The 
above probability therefore becomes
\begin{equation}
\hbox{(probability of $E_0$ given $|a_0\rangle$)}
= \frac{2}{3}\cos^4(\pi/8)
= \frac{1}{4} + \frac{\sqrt{2}}{6} = 0.486.  \label{half}
\end{equation}
The probability of $F_0$ given $|a_0\rangle$ must have this same
value, and the remaining probability must be split equally among
the four remaining outcomes.  Thus, if the initial state is $|a_0\rangle$,
the probabilities of the six outcomes are
\begin{equation}
\frac{1}{4}+\frac{\sqrt{2}}{6},\hspace{3mm}
\frac{1}{4}+\frac{\sqrt{2}}{6},\hspace{3mm}
\frac{1}{8}-\frac{\sqrt{2}}{12},\hspace{3mm}
\frac{1}{8}-\frac{\sqrt{2}}{12},\hspace{3mm}
\frac{1}{8}-\frac{\sqrt{2}}{12},\hspace{3mm}
\frac{1}{8}-\frac{\sqrt{2}}{12}.
\end{equation}
Moreover, because of the symmetry of the problem
we get the same set of values for each of the other
two possible initial states $|a_1\rangle$ and $|a_2\rangle$.
The six measurement outcomes are all equally likely
{\em a priori}; so the mutual information is
\begin{equation}
I = \log 6 + 2\left(\frac{1}{4}+ \frac{\sqrt{2}}{6}\right)
\log\left(\frac{1}{4}+ \frac{\sqrt{2}}{6}\right)
+ 4\left(\frac{1}{8}- \frac{\sqrt{2}}{12}\right)
\log\left(\frac{1}{8}- \frac{\sqrt{2}}{12}\right),
\end{equation}
which comes out to be 1.369 bits.  This is exactly 
what we got with the entangled measurement of Section 2.
Indeed, by comparing Eq.~(\ref{half})
with Eq.~(\ref{whole}), one can already see that the mutual
information for the two cases will be the
same.  (Each probability for the six-outcome measurement
is one-half of the corresponding value for the three-outcome entangled
measurement, but the {\em a priori} probabilities are also
half as great, and this overall factor of one-half has no effect on the mutual
information.)

We have thus found an alternative measurement that is just
as good as our best entangled measurement (at least, the best that
is known so far).  And
this alternative measurement is {\em unentangled}.  The exact equivalence
is perhaps a little surprising, since our construction of 
the unentangled measurement is quite different from that of
the entangled measurement and is motivated by a different heuristic strategy.  
In fact, one might even wonder why the unentangled 
measurement could not have been a little {\em better} than the
entangled one.  

The following section shows that there is, after all, a connection
between the two cases, and that the equivalence between these two 
measurements, regarded as strategies for discriminating the states of the double-trine
ensemble, is not as surprising as it might at first seem.

\section{Assimilating the singlet state}

Recall that the entangled measurement discussed in Section 2 is 
defined by the orthonormal basis $\{|A_0\rangle, |A_1\rangle,
|A_2\rangle, |S\rangle\}$, the last element of which is the
singlet state.  The unentangled measurement of Section 3 is defined
by six non-orthogonal states $\{|B_0\rangle, |B_1\rangle, |B_2\rangle,
|C_0\rangle, |C_1\rangle, |C_2\rangle\}$.  It is useful to write
a few of these states out explicitly in the standard basis
$\{|00\rangle, |01\rangle, |10\rangle, |11\rangle\}$:
\begin{equation}
|A_0\rangle = \frac{1}{\sqrt{6}}\mtx{c}{1\hspace{-1mm}+\hspace{-1mm}\sqrt{2} \\
0 \\ 0 \\ 1\hspace{-1mm}-\hspace{-1mm}\sqrt{2}}\hspace{5mm}
|B_0\rangle = \frac{1}{2\sqrt{2}}\mtx{c}{1\hspace{-1mm}+\hspace{-1mm}\sqrt{2} \\
-1 \\ 1 \\ 1\hspace{-1mm}-\hspace{-1mm}\sqrt{2}} \hspace{5mm}
|C_0\rangle = \frac{1}{2\sqrt{2}}\mtx{c}{1\hspace{-1mm}+\hspace{-1mm}\sqrt{2} \\
1 \\ -1 \\ 1\hspace{-1mm}-\hspace{-1mm}\sqrt{2}}  \label{A0}
\end{equation}
In terms of the singlet state $|S\rangle = (1/\sqrt{2})
(|01\rangle - |10\rangle)$, we can write $|B_0\rangle$
and $|C_0\rangle$ as 
\begin{equation}
|B_0\rangle = \frac{\sqrt{3}}{2}|A_0\rangle - \frac{1}{2}
|S\rangle, \hspace{10mm}
|C_0\rangle = \frac{\sqrt{3}}{2}|A_0\rangle + \frac{1}{2}
|S\rangle .  \label{abc}
\end{equation}

A similar relation holds for the other values of the index
$j$.  To see this, note that the single-qubit
unitary operator
\begin{equation}
u = \mtx{cc}{-1/2 & \sqrt{3}/2 \\ -\sqrt{3}/2 & -1/2}
\end{equation}
maps our three initial single-qubit states to each other
cyclically; that is,
for each $j=0,1,2$, we have
$u|\psi_j\rangle = |\psi_{j+1}\rangle$, where the addition is
mod 3 as usual.  It follows that the tensor product operator
$U = u\otimes u$ generates similar cycles among our two-qubit
states:
\begin{equation}
U|A_j\rangle = |A_{j+1}\rangle, \hspace{5mm}
U|B_j\rangle = |B_{j+1}\rangle, \hspace{5mm}
U|C_j\rangle = |C_{j+1}\rangle, \hspace{5mm}
j = 0, 1, 2.
\end{equation}
Moreover, $U$ leaves the singlet state $|S\rangle$ invariant;
so applying $U$ to each term in Eq.~(\ref{abc}), we get
\begin{equation}
|B_j\rangle = \frac{\sqrt{3}}{2}|A_j\rangle - \frac{1}{2}
|S\rangle, \hspace{7mm}
|C_j\rangle = \frac{\sqrt{3}}{2}|A_j\rangle + \frac{1}{2}
|S\rangle , \hspace{5mm}j = 0, 1, 2.  \label{ABC}
\end{equation}
Thus the states that define our unentangled measurement are
quite closely related to the states of our entangled measurement:
the separable states $|B_j\rangle$ and $|C_j\rangle$ are 
obtained from the entangled state $|A_j\rangle$
by superposition with the singlet.  The singlet, being
orthogonal to all three states $|a_j\rangle$ of our initial ensemble,
does not contribute to the probabilities of any of the
six outcomes of our unentangled measurement.  Moreover,
the fact that the singlet has the same weight in all six
of the POVM elements of this measurement ensures that
its presence will have no effect on the mutual information.
In other words, the effective equivalence between the measurements
of Sections 2 and 3 
follows fairly
directly from Eq.~(\ref{ABC}).

In light of this observation, it is reasonable to ask whether we could
have seen from the outset, without going through the construction in 
Section 3, that {\em some} superpositions
of the form
\begin{equation}
\beta |A_j\rangle - \gamma |S\rangle \hspace{3mm}\hbox{and}
\hspace{3mm}
\beta |A_j\rangle + \gamma |S\rangle  \label{sup}
\end{equation}
would be separable for all $j=0,1,2$.  If so, then we could
have tried to use this fact to construct a suitable POVM. 

In fact we {\em could} have seen this.  A simple measure of the entanglement
of a pure two-qubit state $|v\rangle$ is the concurrence $C$ \cite{HW}, which can be 
written as
\begin{equation}
C = |\langle \bar{v}|\sigma_y\otimes \sigma_y|v\rangle |,
\end{equation}
where $\langle\bar{v}|$ is obtained from $\langle v |$ by 
complex conjugation in the standard basis.  For our present 
purpose it is useful to consider the related quantity
\begin{equation}
c = \langle \bar{v}|\sigma_y\otimes \sigma_y|v\rangle,
\end{equation}
which is $C$ without the absolute value.  Evaluating
$c$ for the superposition
$\beta|A_0\rangle + \gamma|S\rangle$, we have
$$
\hspace{-2.4cm}c = (\beta\langle A_0|+ \gamma\langle S|)
(\sigma_y\otimes\sigma_y) (\beta|A_0\rangle + \gamma| S\rangle) \hfill
$$
\begin{equation}
= \beta^2\langle A_0|\sigma_y\otimes \sigma_y|A_0\rangle
+ \gamma^2\langle S | \sigma_y\otimes \sigma_y| S\rangle \label{plusminus}
= \beta^2/3 - \gamma^2 ,
\end{equation}
where we have used the fact that $\langle A_0|\sigma_y\otimes
\sigma_y|S\rangle = 0$, and in the last step we have
evaluated $\langle A_0|\sigma_y\otimes \sigma_y|A_0\rangle$ starting
from Eq.~(\ref{A0}).  From Eq.~(\ref{plusminus}) it is clear
that we can choose real $\beta$ and $\gamma$ so that $c=0$
for the state 
$\beta|A_0\rangle + \gamma|S\rangle$, and that the
same $\beta$ and $\gamma$ will work for the superposition
$\beta|A_0\rangle - \gamma|S\rangle$.  So these states
can both be made unentangled.  The values
of $\beta$ and $\gamma$ derived in this way are $\beta = \pm \sqrt{3}/2$
and $\gamma = \pm 1/2$, in agreement with
Eq.~(\ref{abc}).  The states given in 
Eq.~(\ref{sup}) with $j=1$ and $j=2$ can similarly be made unentangled, with the 
same values of $\beta$ and $\gamma$, since the unitary operator
$U$, which takes $|A_j\rangle$ to $|A_{j+1}\rangle$
and leaves $|S\rangle$ invariant, 
commutes with $\sigma_y\otimes\sigma_y$.

So far in this section we have focused on the requirement that the states in Eq.~(\ref{sup})
be separable.  But we also want these states to constitute a POVM.
That is, we want
\begin{equation}
\frac{2}{3}\;\sum_{j=0}^2 [
(\beta |A_j\rangle + \gamma |S\rangle) (\bar{\beta}\langle A_j|+\bar{\gamma}\langle S|)
+ 
(\beta |A_j\rangle - \gamma |S\rangle)(\bar{\beta}\langle A_j|-\bar{\gamma}\langle S|)]= I.
\end{equation}
This requirement can be simplified to the form
\begin{equation}
4|\gamma|^2 |S\rangle\langle S|+ (4/3)|\beta|^2\sum_{j=0}^2 |A_j\rangle\langle A_j|
= I.
\end{equation}
Now, we know that 
\begin{equation}
|S\rangle\langle S| + \sum_{j=0}^2 |A_j\rangle\langle A_j| = I.
\end{equation}
So the six states given in Eq.~(\ref{sup}) will constitute a POVM
only if $|\beta|^2 = 3/4$ and $|\gamma|^2 = 1/4$.  If we take $\beta$
and $\gamma$ to be real, we get the same values that we obtained above by requiring that
the states in Eq.~(\ref{sup}) be separable.

We thus have two distinct arguments leading to the conclusion that 
$\beta^2$ should be three times as large as $\gamma^2$ (assuming now that $\beta$
and $\gamma$ are real).  The latter argument---based on the requirement
that the states of Eq.~(\ref{sup}) constitute a POVM---ultimately boils down
to the fact that there are three $|A_j\rangle$'s and only one $|S\rangle$.  
The former argument---based on the requirement that the states of
Eq.~(\ref{sup}) be separable---ultimately boils down to the fact that 
each of the $|A_j\rangle$'s is only 1/3 as entangled as the singlet state,
entanglement being measured by the concurrence.  If one could find some
general principle ensuring that the sum of the concurrences of the 
$|A_j\rangle$'s is equal to the concurrence of $|S\rangle$, then these
two facts could be seen as closely related and the existence of our unentangled and
(apparently) optimal POVM would be less surprising.  However, I am not aware
of any such general principle; so for the time being, I am inclined to think that there is
something special about the double-trine ensemble that makes these two distinct
arguments lead to the same conclusion and thereby allows the construction of an 
optimal measurement that is unentangled.  

\section{Discussion}

This paper has focused on one particular ensemble of states, the double-trine.
When Asher Peres and I considered this case originally, our question was whether
the three states of the ensemble could be distinguished better by a global measurement
than by any sequence of local measurements.  The evidence strongly suggests that 
a global measurement is indeed better.  But among global measurements, one can consider
two kinds: those whose POVM operators are all tensor products, and those that include
at least one entangled operator.  In the present paper I have asked whether, for the 
double-trine ensemble, an unentangled POVM can be as good as the best entangled
measurement (at least the best that is known).  We have seen that a specific unentangled POVM
is indeed just as good, and we have found a couple of ways of constructing it.  
In this sense the double-trine ensemble does not exhibit the strongest kind of
nonlocality that one can imagine for an ensemble of product states.  

This result raises the following question which I mentioned in the Introduction.  
Can {\em every} ensemble of product
states be distinguished just as well by an unentangled measurement as by an
entangled measurement?  In other words, if the states that one is trying
to distinguish are products, is it sufficient to consider measurement outcomes
that are also products?  This question has apparently not been answered in the
literature, though much progress has been made on questions of distinguishability
by local measurements \cite{WSHV,Virmani, G1,WH,CL,TDL,HSSH}.  
It is known that certain sets of orthogonal product states that cannot be
distinguished by a sequence of local measurements {\em can} be distinguished
by separable measurements \cite{UPB2}.  
We also know that 
for some ensembles of unentangled {\em mixed} states
(which are not product states but can be expressed as mixtures of product 
states), there exists no unentangled measurement achieving the optimal mutual 
information \cite{B,DLT}.  However, this fact does not provide an answer to our
current question.  The set of all product states is much smaller
than the set of all unentangled states.  

In the preceding section, we saw, for our specific ensemble, how one could start with an entangled measurement
and use it to construct an {\em unentangled} measurement that provided just as much 
information.  Our method was to superpose, with each of the states representing
a possible outcome of the entangled measurement, a certain proportion of another state (the singlet)
which was orthogonal to each element of the given ensemble.  One can imagine trying
to generalize this construction, but it is not at all clear from this one example how
one might do this.  In any case, it would be good to settle the 
question about the power of unentangled measurements.  
If it turns out that every ensemble of product states {\em can}
be distinguished optimally by an unentangled measurement, then, in a sense,
quantum mechanics is not as nonlocal a theory as one might have imagined.

\end{document}